\documentclass[preprint,floatfix] {revtex4} 
\newcommand{\rvec}{\mathrm {\mathbf {r}}} 
\newcommand{\pvec}{\mathrm {\mathbf {p}}} 
\usepackage{graphicx}
\usepackage{subfigure}
\usepackage{xcolor}
\usepackage{amsmath}
\usepackage{enumerate}
\usepackage{longtable}
\usepackage{tabularx}

\usepackage{color, soul}
\definecolor{darkblue}{rgb}{0,0,0.5}
\setulcolor{darkblue}

\begin{document}

\title{Fisher information in confined hydrogen-like atoms}

\author{Neetik Mukherjee}
%%\altaffiliation{Email: neetik.mukherjee@iiserkol.ac.in.}

\author{Sangita Majumdar}

\author{Amlan K.~Roy}
\altaffiliation{Corresponding author. Email: akroy@iiserkol.ac.in, akroy6k@gmail.com.}
\affiliation{Department of Chemical Sciences\\
Indian Institute of Science Education and Research (IISER) Kolkata, 
Mohanpur-741246, Nadia, WB, India}

\begin{abstract}
%%1234567890 %%1234567890 %%1234567890 %%1234567890 %%1234567890 %%1234567890 %%1234567890 %%1234567890 %%1234567890 %%1234567890
Fisher information ($I$) is investigated for confined hydrogen atom (CHA)-like systems in conjugate $r$ and $p$ spaces. A 
comparative study between CHA and free H atom (with respect to $I$) is pursued. In many aspects, inferences in CHA are 
significantly different from free counterpart; that includes its dependence on $n, l, m$. The role of atomic number and atomic 
radius is discussed. Further, a detailed systematic result of $I$ with respect to variation of confinement 
radius $r_c$ is presented, with particular emphasis on \emph{non-zero}-$(l,m)$ states. Several new interesting observations are 
recorded. Most of these results are of benchmark quality and presented for the first time.  

\vspace{5mm}

\vspace{5mm}

\end{abstract}
\maketitle

\section{introduction}

In first half of the twentieth century, Michels \emph{et al.} \cite{michels37} designed and proposed a simple model in which 
a hydrogen atom was enclosed in an impenetrable spherical cavity keeping the nucleus at centre. For these confined quantum 
systems, the wave function vanishes at a certain boundary which lies at a finite distance, but may be extended up to infinity. 
In such a situation, the particle shows interesting distinctive changes in its observable properties \cite{aquino16,yu17}. 
Such a model of confinement can be exploited as realistic approximation to various physical and chemical environments 
\cite{katriel12}, with special importance in the field of condensed matter, semiconductor physics, astrophysics, 
nano-science and technology, quantum dots, wires and wells \cite{sabin2009,pang11, sen2014electronic}. In last few decades, 
confined quantum systems like atoms, molecules either in fullerene cage or inside the cavities of zeolite molecular sieves, 
in solvent environments etc., have been explored extensively \cite{sabin2009,sen12,sen2014electronic}.          

In recent years, information theory has emerged as a subject of topical interest. At a fundamental level, this explicitly 
deals with single-particle probability density $\rho(\tau)$ of a system. Hence, statistical quantities directly related to 
$\rho(\tau)$ have their importance in predicting and explaining numerous interesting phenomena in both physics as well as in 
chemistry  \cite{frieden04}. A few examples of them are information entropies like R\'enyi ($R$) and Shannon ($S$) entropy, 
Fisher information ($I$), Onicescu energy ($E$), etc. As a consequence of the fact that, $I$ represents the gradient 
functional of density, it measures the local fluctuation of space variable. An increase in $I$ indicates localization of the 
particle. In other words with rise in $I$, the density distribution gets concentrated as well as uncertainty reduces 
\cite{frieden04}. It is important to note that, $I$ resembles the Weizs\"acker kinetic energy functional
($T_{\omega}[\rho]$) frequently used in density functional theory (DFT) \cite{debajit12}. Lately, an \emph{Euler} equation
in orbital-free DFT has been formulated with the help of $I$ and $S$ \cite{nagy15}. For spherically symmetric 
systems this equation can be formalized by using only a specific form of $I$ \cite{nagy15}. Because of its ability to predict 
and explain versatile properties, $I$ has been especially invoked to explore Pauli effects \cite{nagy06a,toranzo14}, 
ionization potential, polarizability \cite{sen07}, entanglement \cite{nagy06b}, avoided crossing \cite{ferez05} etc., in 
atomic systems. In molecular systems, $I$ has been exploited to investigate steric effect \cite{nagy07,esquivel11}, bond 
formations \cite{nalewajski08}, elementary chemical reactions \cite{rosa10} etc.       

About a decade ago, numerical investigation of $I$ for ground-state of neutral atoms \cite{romera02,romera04} was made. Some 
analysis from analytical standpoint was given in \cite{romera05}, where the authors formulated a pair of equations to compute 
$I_{\rvec}$, $I_{\pvec}$ in \emph{central potentials}. Accordingly, they are expressed in terms of 
four expectation values \emph{viz.,} $\langle p^2\rangle,~\langle r^{-2}\rangle$ and 
$\langle r^2\rangle,~\langle p^{-2}\rangle$ respectively. In recent time, $I$ in both $r$ and $p$ spaces have been reported
for various model diatomic potentials, such as P\"oschl-Teller \cite{dehesa06}, pseudo-harmonic \cite{yahya14}, Tietz-Wei 
\cite{falaye14}, Frost-Musulin \cite{idiodi16}, Generalized Morse \cite{onate16}, exponential-cosine screened coulomb 
\cite{abdelmonem17} potential, etc.

Study of $I$ in a \emph{confined} hydrogen atom (CHA) are quite scarce. We are aware of only the work of \cite{aquino13}, 
where it was considered for CHA under soft and hard confinement \cite{aquino13} for ground state only. In this endeavour, 
our primary objective is to perform an explicit analysis of $I$ in a CHA-like system, for any arbitrary state characterized 
by principal, azimuthal and magnetic quantum numbers $n, l, m$ in conjugate spaces with special emphasis on $m \neq 0$ states. 
Elucidative calculations are performed with exact analytical wave functions in $r$-space; whereas the $p$-space, wave 
functions are obtained from numerical Fourier transform of $r$-space counterpart. Representative results are given for
$2p, 3d, 4f, 5g$ as well as $10s-10m$ states, to understand the various effects. Here, we have envisaged all the allowed 
$m$'s corresponding to a given $n$ and $l$, which allows one to follow the detailed changes in behavior of states with 
different $m$ as the environment switches from free to confinement. Changes are also monitored with respect to $Z$ in  
H-isoelectronic series under confinement. Since such works are very limited, most of the 
current results are provided here for the first time. Section~II gives a brief description about the theoretical method 
used; Sec.~III offers a detailed discussion of results of $I$, while we conclude with a few comments in Sec.~IV.

\section {Methodology}
The non-relativistic radial Schr\"odinger equation for a confined H-like atom, without any loss of generality, may be written
as (atomic unit employed, unless otherwise mentioned), 
\begin{equation}
\left[-\frac{1}{2} \ \frac{d^2}{dr^2} + \frac{\ell (\ell+1)} {2r^2} + v(r) +v_c (r) \right] \psi_{n,\ell}(r)=E_{n,\ell}\ 
\psi_{n,\ell}(r),
\end{equation}
where $v(r)=-Z/r \ (Z=1$ for H atom). Our required confinement inside an impenetrable spherical cage is introduced by 
invoking the following form of potential: $v_c(r) = +\infty$ for $r > r_c$, and $0$ for $r \leq r_c$, where $r_c$ implies 
radius of confinement. 

\emph{Exact} generalized radial wave function for a CHA is mathematically expressed as \cite{burrows06}, 
\begin{equation}
\psi_{n, l}(r)= N_{n, l}\left(2r\sqrt{-2\mathcal{E}_{n,l}}\right)^{l} \ _{1}F_{1}
\left[\left(l+1-\frac{1}{\sqrt{-2\mathcal{E}_{n,l}}}\right),(2l+2),2r\sqrt{-2\mathcal{E}_{n,l}}\right] 
e^{-r\sqrt{-2\mathcal{E}_{n,l}}},  
\end{equation}
where $N_{n, l}$ represents normalization constant and $\mathcal{E}_{n,l}$ denotes energy eigenvalue of a given state 
distinguished by $n,l$ quantum numbers, whereas $_1F_1\left[a,b,r\right]$ is a confluent hypergeometric function. Allowed 
energies are enumerated by imposing Dirichlet boundary condition, $\psi_{n,l} (0)=\psi_{n,l}(r_c)=0$ in Eq.~(2). In this work,
generalized pseudospectral (GPS) method has been applied to compute $\mathcal{E}_{n,l}$ of CHA. This method has produced  
very accurate results for various model and real systems including atoms and molecules in the last decade; some of which 
could be found in the references \cite{roy04,sen06,roy13, roy15}.  

The $p$-space wave function is obtained numerically from Fourier transform of $r$-space counterpart, and as such given as, 
\begin{equation}
\begin{aligned}
\psi_{n,l}(p) & = & \frac{1}{(2\pi)^{\frac{3}{2}}} \  \int_0^\infty \int_0^\pi \int_0^{2\pi} \psi_{n,l}(r) \ 
\Theta_{l,m}(\theta) \Phi_{m}(\phi) \ e^{ipr \cos \theta}  r^2 \sin \theta \ \mathrm{d}r \mathrm{d} \theta \mathrm{d} \phi .
\end{aligned}
\end{equation}
Here $\psi(p)$ needs to be normalized. The normalized $r$- and $p$-space densities are represented as, 
$\rho(\rvec) = |\psi_{n,l,m}(\rvec)|^2$ and $\Pi(\pvec) = |\psi_{n,l,m} (\pvec)|^2$ respectively. 
Let $I_{\rvec}$, $I_{\pvec}$ denote \emph{net} information measures in conjugate $r$ and $p$ space of CHA. It is well
established that, for a single particle in a central potential, these quantities can be written in terms of radial 
expectation values $\langle r^k \rangle $ and $ \langle p^k \rangle, (k = -2,2)$ \cite{romera05}, as below, 
\begin{eqnarray} 
I_{\rvec}  =  \int_{{\mathcal{R}}^3} \left[\frac{|\nabla\rho(\rvec)|^2}{\rho(\rvec)}\right] \mathrm{d}\rvec  =  
4\langle p^2\rangle - 2(2l+1)|m|\langle r^{-2}\rangle \\ 
I_{\pvec} =  \int_{{\mathcal{R}}^3} \left[\frac{|\nabla\Pi(\pvec)|^2}{\Pi(\pvec)}\right] \mathrm{d} \pvec  = 
4\langle r^2\rangle - 2(2l+1)|m|\langle p^{-2}\rangle.
\end{eqnarray}
The above equations can be further recast in the following forms,
\begin{eqnarray}
I_{\rvec} = 8\mathcal{E}_{n,l}-8\langle v(r)\rangle-2(2l+1)|m|\langle r^{-2}\rangle \\
I_{\pvec} = 8\mathcal{E}_{n,l}-8\langle v(p)\rangle-2(2l+1)|m|\langle p^{-2}\rangle .
\end{eqnarray}
where $v(p)$ is the $p$-space counterpart of $v(r)$. 

In case of H-isoelectronic series, $I$'s in $r$ and $p$ space are expressed as;
\begin{equation}
\begin{aligned} 
I_{\rvec}(Z)  =Z^{2}I_{\rvec}(Z=1), \hspace{3mm} \ \ \ \  I_{\pvec}(Z) = \frac{1}{Z^{2}}I_{\pvec}(Z=1).
\end{aligned}
\end{equation}
Hence, an increase in $Z$ leads to rise in $I_{\rvec}(Z)$ and fall in $I_{\pvec}(Z)$. However, it is obvious that 
$I_{t} \ (=I_{\rvec} I_{\pvec})$ remains invariant with $Z$. Throughout this work, $I_{\rvec}(Z=1)$ and $I_{\pvec}(Z=1)$ will 
be denoted as $I_{\rvec}$, $I_{\pvec}$ respectively.

When $m=0$, $I_{\rvec}$ and $I_{\pvec}$ in Eqs.~(4), (5) reduce to simplified forms as below,  
\begin{equation}
\begin{aligned} 
I_{\rvec}  =  4\langle p^2\rangle, \hspace{3mm} \ \ \ \  I_{\pvec} =4\langle r^2\rangle.
\end{aligned}
\end{equation}
It is seen that, at a fixed $n, l$, both $I_{\rvec}$ and $I_{\pvec}$ are maximum when $m=0$, decreasing with rise in $m$. 
Hence one obtains the following upper bound for $I_t$, 
\begin{equation}
 I_{\rvec} I_{\pvec} \ (=I_t) \leq 16 \langle r^2\rangle \langle p^2\rangle
\end{equation}
Further manipulation using Eqs.~(6) and (7) leads to following uncertainty relations \cite{romera05}, 
\begin{equation}
\frac{81}{\langle r^2\rangle \langle p^2\rangle} \leq I_{\rvec} I_{\pvec} \leq 16 \langle r^2\rangle \langle p^2\rangle.  
\end{equation}
Therefore, in a central potential, $I$-based uncertainty product is bounded by both upper and lower limits. They
are state dependent, varying with alterations in $n, l, m$. 

\begingroup            %Table~I  (2p-state)
\squeezetable
\begin{table}
\caption{$I_{\rvec}, I_{\pvec}$ and $I_{t}$ for $2p$ orbitals at seven different $r_c$. See text for details.}
\centering
\begin{ruledtabular}
\begin{tabular}{l|lllllll}
 $|m|$ &  $r_c=0.1$ & $r_c=0.3$  & $r_c=0.5$   &   $r_c=1$ &   $r_c=2.5$   &   $r_c=5$  &   $r_c=10$   \\
\hline
\multicolumn{8}{c}{$I_{\rvec}$}  \\
\hline
 0   &   8076.456640  &   897.5338363 & 323.2227616  & 80.94182631 & 13.1277640 & 3.494588403  & 1.2520908545 \\
 1   &   5724.072262 &    633.2807489 & 227.02170314  & 56.182977183 & 8.762099224 & 2.160171627  & 0.6653371253 \\
\hline
\multicolumn{8}{c}{$I_{\pvec}$}  \\
\hline
 0   &   0.0149412219  &  0.1336873908 & 0.3691433181  & 1.4538574615 & 8.6256520961 & 30.9047630304   & 87.12258908648 \\
 1   &   0.009996    &  0.08930        & 0.2462       & 0.9657    & 5.65    & 19.74     & 51.92   \\
\hline
\multicolumn{8}{c}{$I_{t}^{\ddag}$}  \\
\hline
 $0^{\dag}$ &   120.672131 & 119.9889567 &  119.3155227  &  117.67787813 &  113.2355250  &  107.9994264835 & 109.0853970155  \\
 1          &   57.217826  & 56.55197    &  55.8927      &  54.2559 &  49.506  &  42.64 & 34.54  \\
\end{tabular}
\end{ruledtabular}
\begin{tabbing}
$^\dag$These also correspond to upper bounds, given in Eq.~(10). \\
$^\ddag$Lower bounds, Eq.~(11), at 7 $r_c$ are: 10.739845, 10.8009939, 10.8619563, 11.01311496, 11.4451714, 12.00006372, 11.8806003.
\end{tabbing}
\end{table}
\endgroup

\section{Result and Discussion}
At first, it is necessary to mention a few points for ease of discussion. Effect of quantum number $m$ on $I_{\rvec}$ and 
$I_{\pvec}$ of CHA is the main focus of our work; which is attempted here for first time. Therefore, to ensure a good 
accuracy of calculated quantities, a series of tests were performed. The results in various tables are presented up to  
those points which maintained convergence. $I_{\rvec}$ values are obtained from Eqs.~(4) and (6), whereas $I_{\pvec}$ from 
Eq.~(5). In all occasions, it has been verified that, in both spaces, as $r_c \rightarrow \infty$, $I_{\rvec}$ and 
$I_{\pvec}$ coalesce to respective free Hydrogen atom (FHA) limit. Here, \emph{net} $I$ in $r$ and $p$ spaces are segmented 
into radial and angular part. But in both $I_{\rvec}$ and $I_{\pvec}$ expressions, angular part is normalized to unity. 
Hence, evaluation of all these desired quantities using only radial part will serve the purpose. The radial parts of 
wave function in $r$ and $p$ spaces depend on $n, l$ quantum numbers. Hence, $p$-space radial wave function can be generated
by putting $m=0$ in Eq.~(3). Further, a change in $m$ from \emph{zero} to \emph{non-zero} value will not affect the 
form of radial wave function in $p$ space. Confinement in hydrogen atom is achieved by pressing the radial boundary 
from infinity to a finite region. To achieve these, pilot calculations are done for $2p,~3d,~4f,~5g$ and all $l$-states
corresponding to $n=10$, varying $r_c$ from 0.1 to 100 a.u. We chose $2p,~3d,~4f,~5g$ as they represent nodeless ground 
state of a particular $l$, whereas, $10s$-$10m$ states are undertaken to understand the effect of nodes on $I$ at 
\emph{non-zero} $m$.

\begingroup            %Table~II  (3d-state)
\squeezetable
\begin{table}
\caption{$I_{\rvec}, I_{\pvec}$ and $I_{t}$ for $3d$ orbitals at seven selected $r_c$. See text for details.}
\centering
\begin{ruledtabular}
\begin{tabular}{l|lllllll}
 $|m|$ &  $r_c=0.1$ & $r_c=0.3$  & $r_c=0.5$   &   $r_c=1$ &   $r_c=2.5$   &   $r_c=5$  &   $r_c=10$   \\
\hline
\multicolumn{8}{c}{$I_{\rvec}$}  \\
\hline
 0   &   13287.04524 & 1476.392686  & 531.5410116   & 132.932946 & 21.327058 & 5.39200335  & 1.43124171 \\
 1   &   10413.81694 & 1155.77393   & 415.6163877   & 103.62877  & 16.469367 & 4.0931882   & 1.0451648 \\
 2   &   7540.588647 & 835.1551838  & 299.6917637   & 74.324598  & 11.611677 & 2.7943730   & 0.6590879 \\ 
\hline 
\multicolumn{8}{c}{$I_{\pvec}$}  \\
\hline
 0   &   0.01752499 & 0.15730840 & 0.43580006 & 1.73133439 & 10.587743 &   40.640796     &   146.06896 \\
 1   &   0.01331281 & 0.1194437 &  0.3307477  &   1.312436 &   7.99650 &   30.48999      &   107.87459 \\
 2   &   0.00910063 & 0.0815791 &  0.225695   &   0.89353  &   5.40526 &   20.33919      &   69.6802   \\
\hline
\multicolumn{8}{c}{$I_{t}^{\ddag}$}  \\
\hline
$0^{\dag}$  &   232.85533  & 232.248971  &  231.6456047   & 230.1513809  & 225.805409  &  219.1353081  & 209.0599880   \\
 1          &   137.63716  & 138.049914  &  137.4641643   & 136.0061283  & 131.697293  &  124.8012672  & 112.7467242   \\
 2          &   67.624107  & 68.1312082  &  67.63893260   & 66.41125805  & 62.7641332  &  56.83528337  & 45.92537668   \\
\end{tabular}
\end{ruledtabular}
\begin{tabbing}
$^\dag$These also correspond to upper bounds, given in Eq.~(10). \\
$^\ddag$Lower bounds, Eq.~(11), at 7 $r_c$ are: 5.56568, 5.580218, 5.5947532, 5.6310763, 5.739455, 5.9141541, 6.1991776. 
\end{tabbing}
\end{table}
\endgroup

The change of probability density in chemical or physical environment can be quantified by $I$. Exact analytical form of 
$I_{\rvec}$ and $I_{\pvec}$ in free H-like atom were given in \cite{romera05},
\begin{equation}
I_{\rvec}(Z)=\frac{4Z^2}{n^2}\left[1-\frac{|m|}{n}\right], \ \ \ I_{\pvec}(Z)=\frac{2n^2}{Z^2} 
          \left[ \left(5n^2+1-3l(l+1)\right)-|m|(8n-6l-3) \right]. 
\end{equation}
It suggests that, at a fixed $m$, $I_{\rvec}(Z)$ diminishes as $n$ goes high, and for a given $n$, lowers as $|m|$ is raised.
But it remains unaffected with changes in $l$. On the other hand, at a particular $l,m$, $I_{\pvec}(Z)$ advances with $n$; but 
reduces with progress of $l$ for constant $n,m$ values. Similarly, for a specific $n, l$, $I_{\pvec}(Z)$ abates with growth 
in $|m|$. By putting $Z=1$ in Eq.~(12) one easily gets the expressions for $I_{\rvec}$ and $I_{\pvec}$ in a FHA. 
  
\begin{figure}                         %%%Fig. 1, CHA
\begin{minipage}[c]{0.35\textwidth}\centering
\includegraphics[scale=0.50]{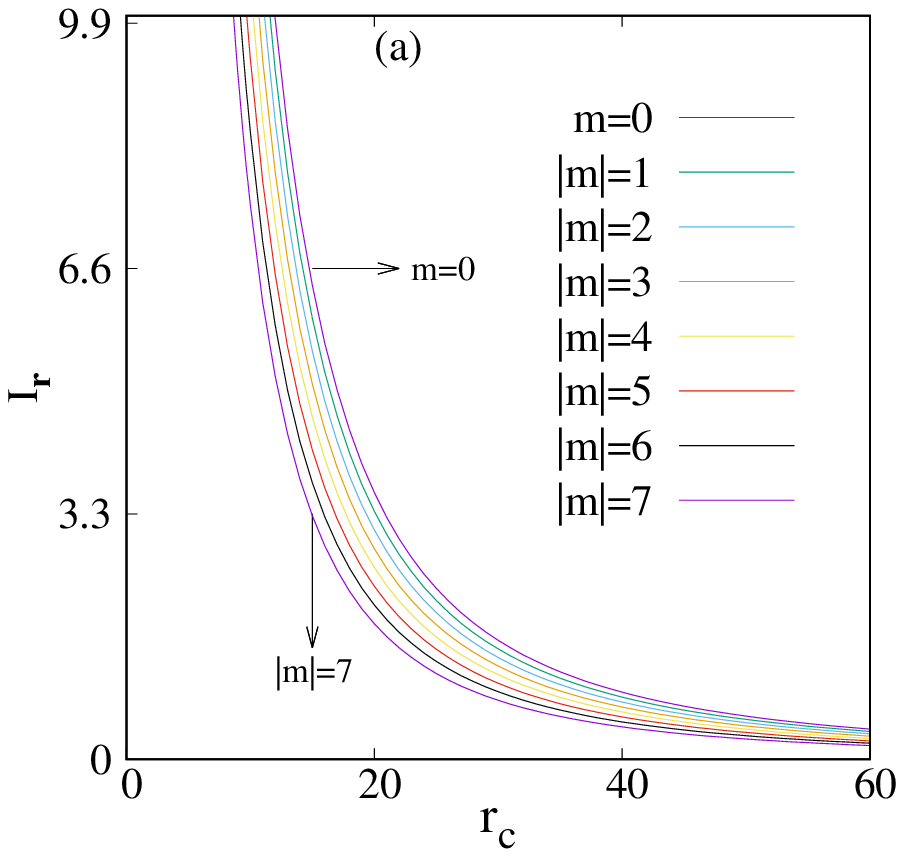}
\end{minipage}%
\vspace{1mm}
\begin{minipage}[c]{0.35\textwidth}\centering
\includegraphics[scale=0.50]{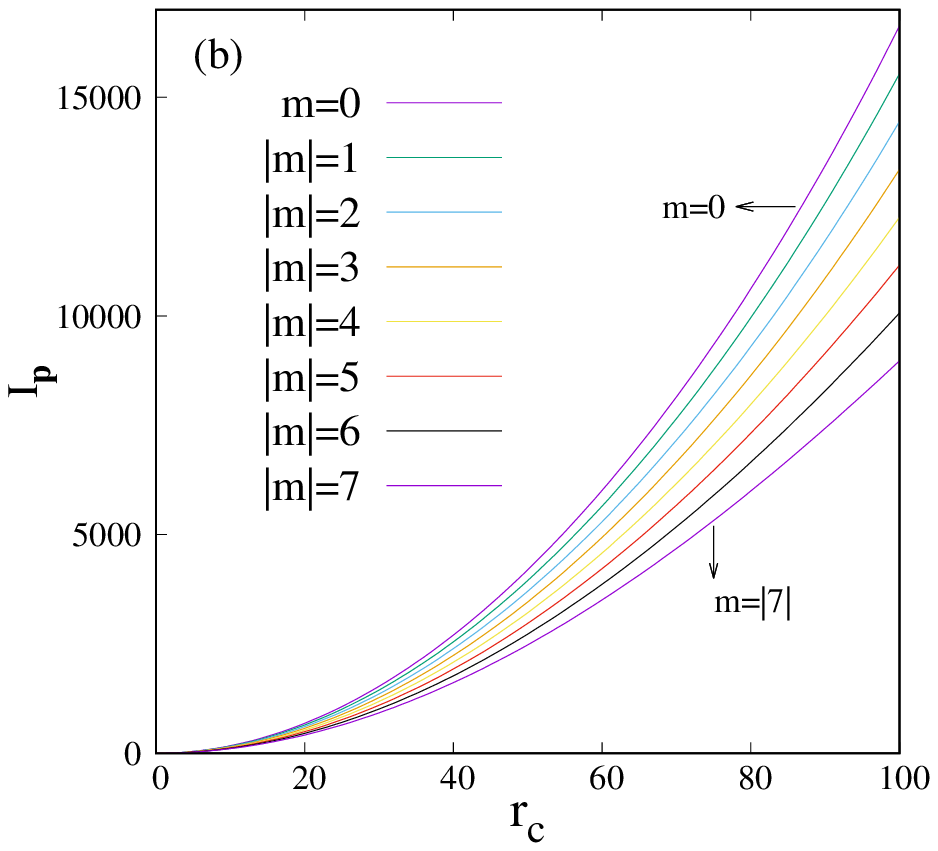}
\end{minipage}%
\caption{Variation of $I_{\rvec}$ and $I_{\pvec}$ in CHA, with $r_c$, for all allowed $|m|$ values of $10k$ orbital, in panels 
(a) and (b) respectively. See text for details.}
\end{figure}

At first, Tables~I and II present $I_{\rvec}, I_{\pvec}$ as well as the \emph{total} quantity $I_t$ for lowest two nodeless
states having $l \neq 0$, i.e., $2p$ and $3d$ respectively; these are provided for all allowed $|m|$ at 7 carefully selected 
$r_c$, in order to get a clear picture of the effect of $m$ quantum number. To minimize space, similar results for $4f$, $5g$
as well as several states corresponding to $n=10$, \emph{viz.}, $10h, 10k$ and $10m$ are offered in Tables~S1-S5 of 
Supplementary Material (SM), respectively. One notices that, behavior of $I_{\rvec}$ and $I_{\pvec}$ in 
CHA is not always in consonance with FHA; a careful analysis reveals considerable deviation in the patterns. In general, 
$I_{\rvec}$ decreases while $I_{\pvec}$ increases as $r_c$ advances; this is found to hold good for all $|m|$. This is to be
expected, as progression in $r_c$ promotes delocalization in $r$ space and localization in $p$ space. At a certain 
$r_c$ and $m$, $I_{\rvec}$ grows with $n$, this trend is completely opposite to that observed from Eq.~(12) in FHA. Because, 
in CHA, kinetic energy gains with $n$, whereas in FHA it falls off. However, at a fixed $n$, in both CHA and FHA, $I_{\rvec}$ 
lessens as one descends down the table (hike in $|m|$). In contrast to $I_{\rvec}$, $I_{\pvec}$ for all four states 
considered, in both CHA and FHA, portray similar pattern. Thus, at a given $m$, it progresses with $n$; further like 
$I_{\rvec}$, $I_{\pvec}$ also falls with advancement of $m$ at a definite $n$. Usually, $I_{t}$ seems to show a propensity 
towards $I_{\rvec}$ in its behavioral pattern, but not always. In all instances, it satisfies the upper and lower bounds
dictated by Eq.~(11). These variations in $I_{\rvec}, I_{\pvec}$ with $r_c$ are recorded for a representative states ($10k$), 
in panels (a), (b) of Fig.~1, for all possible values of $|m|$; the total quantity $I_t$ is produced separately in Fig.~S1 
of SM. For an
arbitrary state characterized by quantum numbers $n,l$, changes in $I_{\rvec}, I_{\pvec}, I_t$ with $r_c$, in a CHA preserve 
same qualitative orderings for various $m$, as in these figures. This has been verified in several other occasions, which are 
not presented here to save space. As usual, in all cases, they all converge to their respective limiting FHA values at some 
sufficiently large $r_c$, which varies from state to state. 

\begingroup            %Table~3 
\squeezetable
\begin{table}[tbp]
\caption{$I_{\rvec}, I_{\pvec}$ for all allowed $l$, corresponding to $n=10$ and $|m|=1$ orbitals in CHA, at seven selected  
$r_c$ values. See text for details.}
\centering
\begin{ruledtabular}
\begin{tabular}{l|lllllllll}
 $l$ &  $r_c=0.1$ & $r_c=0.3$  & $r_c=0.5$  &   $r_c=1$ &   $r_c=2.5$   &   $r_c=5$ &   $r_c=10$   \\
\hline
\multicolumn{8}{c}{$I_{\rvec}$}  \\
\hline
 1   & 337317.31464  & 37474.481640   & 13488.9415784 & 3371.07344973 & 538.824417709 & 134.48580714 & 33.51660547  \\
 2   & 300596.10217  & 33396.594443   & 12021.7096534 & 3004.7684353  & 480.45505152  & 119.99220841 & 29.942974553  \\
 3   & 265034.1043   & 29446.2744769  & 10599.9580236 & 2649.5565064  & 423.728010198 & 105.8538366  & 26.42943372  \\
 4   & 230616.75756  & 25622.70654224 & 9223.6821098  & 2305.6169240  & 368.75853346  & 92.13585342  & 23.0113835   \\
 5   & 197308.08284  & 21922.12562    & 7891.610144   & 1972.6839051  & 315.529017    & 78.8442246   & 19.6956519   \\ 
 6   & 165034.41942  & 18336.439361   & 6600.861984   &  1650.058034  & 263.937368    & 65.95746881  & 16.478900    \\
 7   & 133640.58787  &  14848.450065  & 5345.2624260  & 1336.2054367  & 213.7428638   & 53.4173182   & 13.34749576  \\
 8   & 102757.598457 & 11417.184564   & 4110.0703296  & 1027.4465296  & 164.35944397  & 41.07832028  & 10.26556627  \\
 9   & 71218.59722   & 7913.0133040   & 2848.6264996  & 712.1210846   & 113.92355287  & 28.47535810  & 7.11712852   \\
\hline
\multicolumn{8}{c}{$I_{\pvec}$}  \\
\hline
 1   & 0.0132601      & 0.119372       & 0.33167       & 1.32758      & 8.3142      & 33.374      & 134.520       \\
 2   & 0.013334381    & 0.12002836     & 0.3334650     & 1.3343946    & 8.350289    & 33.47399    & 134.53947     \\
 3   & 0.01350503439  & 0.1215547591   & 0.337678596   & 1.35098345   & 8.44891802  & 33.8338841  & 135.688696    \\
 4   & 0.013813106568 & 0.124318744963 & 0.34533221024 & 1.3813556408 & 8.634146939 & 34.54342445 & 138.265110370 \\
 5   & 0.0143240903   & 0.1289089301   & 0.358058593   & 1.432018965  & 8.94620270  & 35.76039607 & 142.8731406   \\ 
 6   & 0.0151493038   & 0.136326672    & 0.378637879   & 1.51407965   & 9.45422375  & 37.7594223  & 150.5945315   \\
 7   & 0.01649686198  & 0.148444746    & 0.412271503   & 1.648336126  & 10.28805581 & 41.05882556 & 163.492606    \\
 8   & 0.018820152    & 0.1693440515   & 0.470296419   & 1.88014732   & 11.7313926  & 46.7946037  & 186.119273    \\
 9   & 0.02345226     & 0.2110273      & 0.5860667     & 2.343062     & 14.62137    & 58.33161    & 232.06115     \\
\end{tabular}
\end{ruledtabular}
\end{table}
\endgroup

Now to understand the effect of $l$ on $I_{\rvec}, I_{\pvec}$, we offer Table~III, where these are calculated for all the $l$
states having $|m|=1$, for $n$ corresponding to 10, at same selected $r_c$'s of previous tables. Recall that in a FHA, 
$I_{\rvec}$ is independent of $l$. However, unlike FHA, $I_{\rvec}$, $I_{\pvec}$ and $I_{t}$ all depend on $l$ in a CHA; at 
a given $n,m$, the former drops down as $l$ mounts up. This may occur presumably because that, although radial nodes reduce 
with $l$, the orbital density gets more and more diffused. Therefore, at a fixed $n$ and $m$, an orbital with higher $l$ value 
experiences lesser nuclear charge. Whereas $I_{\pvec}$, in a similar occasion, escalates with elevation in $l$ (again for fixed 
$n,m$).

\begin{figure}                         %%%Fig. 2, CHA
\begin{minipage}[c]{0.30\textwidth}\centering
\includegraphics[scale=0.45]{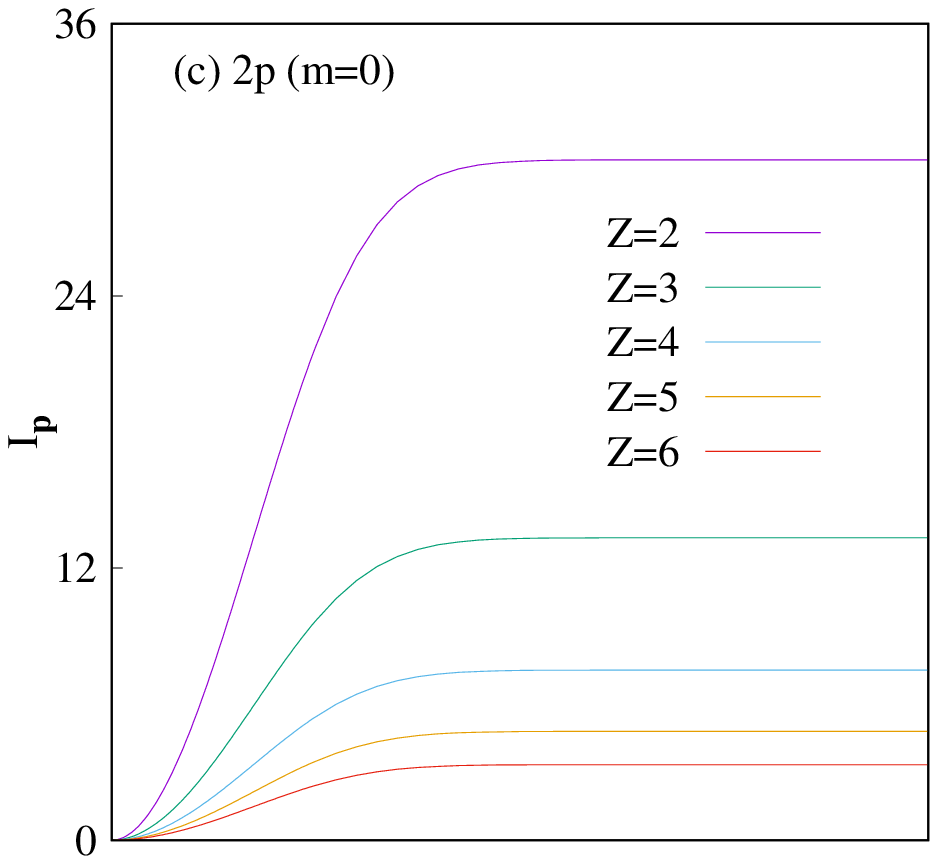}
\end{minipage}%
\vspace{1mm}
\begin{minipage}[c]{0.30\textwidth}\centering
\includegraphics[scale=0.45]{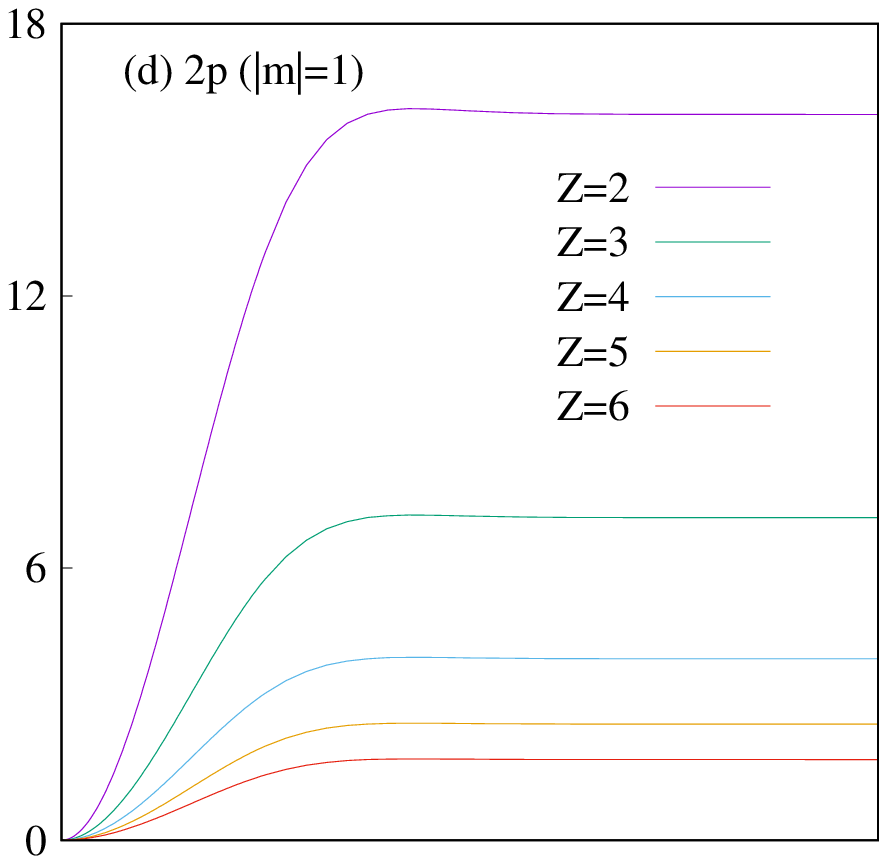}
\end{minipage}%
\vspace{1mm}
\hspace{0.2in}
\begin{minipage}[c]{0.32\textwidth}\centering
\includegraphics[scale=0.48]{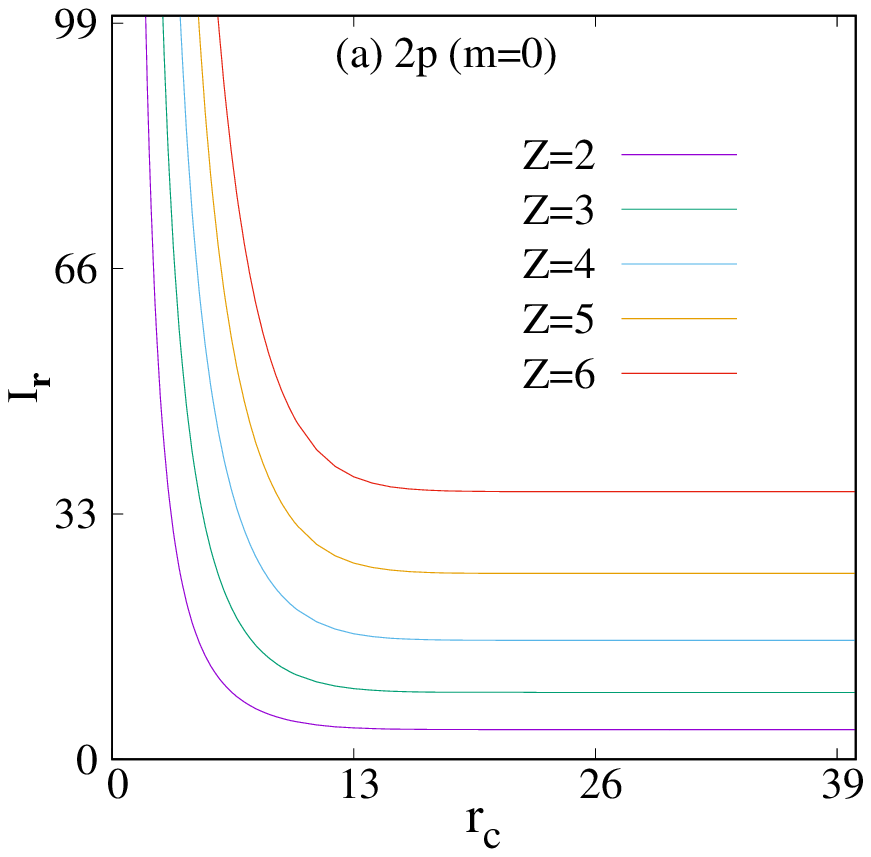}
\end{minipage}%
\begin{minipage}[c]{0.32\textwidth}\centering
\includegraphics[scale=0.48]{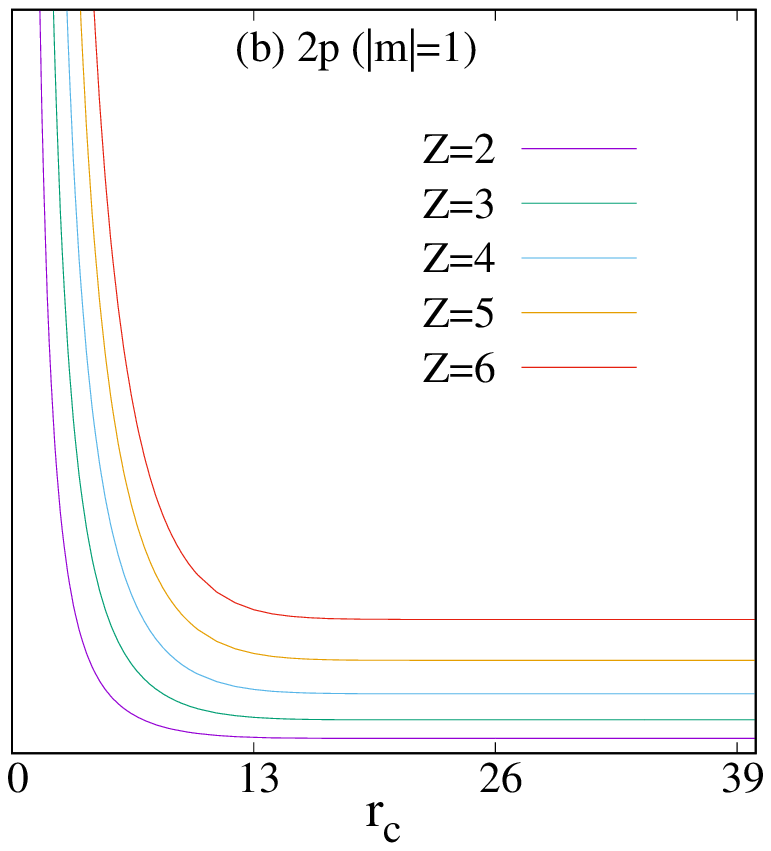}
\end{minipage}%
\caption{Plots of $I_{\rvec}$ (bottom panels (a),(b)) and $I_{\pvec}$ (top panels (c),(d)) in CHA with $r_c$, of 
$2p$ orbital at five selected $Z (2-6)$. Left, right columns correspond to $|m|=0$, 1. See text for details.}
\end{figure}

%%Finally, we study the influence of $l$ on $I$ in CHA. Tables~S3, S4 represent $I_{\rvec}, I_{\pvec}$ and $I_{t}$ for $10h$, $10i$ states 
%%successively. Whereas, tables~III and S5 depict $I_{\rvec}, I_{\pvec}$ and $I_{t}$ of $10m$ states respectively. From Tables~S3, S4 and III
%%it has been found that, at an unambiguous $n,m$ set, diminution of $I_{\rvec}$ and improvement of $I_{\pvec}$ occur with gain in $l$. This, 
%%deduction is again in complete disagreement with FHA. Since, in FHA, as we have discussed before, $I_{\rvec}$ remains unaffected and
%%$I_{\pvec}$ reduces with variation of $l$ (at fixed $n$). Further, akin to $2p-5f$ orbitals, at a certain $l$ value, both $I_{\rvec}$ 
%%and $I_{\pvec}$ decrement with swelling of $m$. Now, the focus is on $I_{t}$. For CHA, it also gets subsided with emergence of $l$. Here 
%%also $I_{t}$ obeys the required bound given in Eq.~(11).      

Before concluding, a few words may be devoted to the influence of $Z$ on $I_{\rvec}, I_{\pvec}$. Thus Fig.~2 depicts plots of 
$I_{\rvec}$ and $I_{\pvec}$ against $r_c$, for 5 selected $Z$ (2--6), in bottom panels (a), (b) and top panels (c), (d). These 
are given for $2p$ orbital; left and right panels characterize $|m|=0$ and 1 respectively. Clearly, at a given $r_c$, 
$I_{\rvec}$ grows and $I_{\pvec}$ decays with increment in $Z$. At a fixed $Z$, dependence of these two measures on $n,l,m$
is similar to that in CHA. As $Z$ goes up, there is more localization, hence compactness in electron density enhances as 
one moves to heavier atoms. Thus, $I_{\rvec}$ should increase with reduction of atomic radius. This emphasizes the 
ability of Fisher information to explain various observed chemical phenomena like electronegativity, ionization potential, 
hard-soft interaction etc., in atomic systems.

\section{Future and Outlook}
Benchmark values of $I_{\rvec}, I_{\pvec}, I_t$ are offered in a CHA, with special emphasis on non-zero-$(l,m)$ states. Representative
results are provided for $2p, 3d, 4f, 5g$ as well as $10l$ states. Effect of $m$ on these measures was analyzed in detail. To 
the best of our knowledge, such an analysis in CHA is pursued here for first time. With progress of $r_c$, $I_{\rvec}$ falls 
and $I_{\pvec}$ rises. These are compared and contrasted with FHA results; in many aspects they show different trends. In FHA, 
$I_{\rvec}$ does not depend on $l$. But, in CHA it reduces with $l$ (for fixed $n, m$). Similarly, with growth of $l$ 
(for fixed $n,m$), $I_{\pvec}$ progresses in CHA and declines in FHA. Further, their changes with respect to $Z$ is also 
considered. This study reveals that, $I$ can be successfully used to predict chemical phenomena in atomic and molecular system. 
Therefore, further inspection of $I$ in many-electron systems would be worthwhile and desirable. 

\section{Acknowledgement}
Financial support from DST SERB, New Delhi, India (sanction order: EMR/2014/000838) is gratefully acknowledged. NM thanks DST SERB, New Delhi, India, 
for a National-post-doctoral fellowship (sanction order: PDF/2016/000014/CS). SM is obliged to IISER-K for her JRF fellowship.

\end{document}